\documentstyle[11pt,newpasp,twoside,epsf]{article} 
\def\edcomment#1{\iffalse\marginpar{\raggedright\sl#1\/}\else\relax\fi} 
\reversemarginpar 

\begin{document} 
\title{Preliminary results from VLT (FORS1) photometry of NGC\,6397}
\author{Andreuzzi G., Testa, V.} 
\affil{INAF-OAR, via Frascati 33,00040 Monteporzio Catone (Italy)}
\author{Marconi G.} 
\affil{ESO-Chile, Casilla 19001, Santiago (Chile)} 
\author{Alcaino G., Alvarado F.} 
\affil{Isaac Newton Institute, Casilla 8-9, Correo 9, Santiago (Chile) } 
\author{Buonanno R.}
\affil{University of Torvergata, 00100 Roma (Italy)}
\begin{abstract} 
We present VLT (FORS1) photometry of the lower main sequence (MS)
of the Galactic Globular Cluster (GGC) NGC\,6397, for stars located in
2 fields extending from a region near the cluster center out
to $\simeq$ 10'.	
The obtained CMD shows a narrow MS extending down to
V $\simeq$ 27 (figure c), much deeper than any previous ground based
study and comparable with previous HST photometry (Cool et al. 1996).
The comparison between observed MS Luminosity Functions (LFs)
derived for 2 annuli at different radial distance from the center of
the cluster shows a clear-cut correlation between their slope before 
reaching the turn-over, and the radial position of the observed fields 
inside the cluster area: the LFs become flatter with decreasing radius, a trend
that is consistent with the interpretation of
NGC\,6397 as a dynamically relaxed system.
\end{abstract}
\section{Introduction} 
GC stars fainter than the MSTO are 
substantially unevolved and are located along the MS according to
their initial mass. Their luminosity distribution is usually described
by the Initial Mass Function as modified by subsequent dynamical
evolution to become the Present Day Mass Function.
The mass segregation then should be seen as a radial dependence of the LFs.
but to verify these theoretical predictions we need to obtain a
statistically significative sample of data starting from the central
regions of the cluster. Now this is possible also by ground
observations thank's to the great power of the VLT and the FORS cameras.
\section{Data analysis and discussion} 
Our observations consist of severals 600 s exposures in the
filter V and I Bessel (proposals 63.H-0721(A); 65.H-0531(B)) of two fields
(6.8' $\times$ 6.8') partially overlapping and 
located at $\simeq$ 7.5' southeast of the center of the
cluster. We have used HST observations (Cool et al., 1996 ApJ, 468, 655)
in common with our fields (figure b) to make the calibration of the data.
The LFs (in bin of 0.5 mag V) for the stars in the regions
R$_1$ $<$ 5.5' and 5.5' $\leq$ R$_2$ $\leq$ 9.8' have
been obtained by following the usual procedure and taking into account for the 
incompleteness of the images (figure a). 
It is evident from figure a) that both the observed LFs show a
maximum at V $\simeq$ 22, then decrease steadily to fainter 
luminosities maintaining a similar shape, but a tangible different slope.
In particular the LFR2 (19.5 $\leq$ V $\leq$ 22),
is steeper (x = 0.16 $\pm$ 0.01) than the LFR1 (x = 0.076 $\pm$ 0.003), 
suggesting that the relative ratio luminous stars/faint
stars is greater within S1 than S2 and then confirming the
mass-segregation effect in this cluster.
\footnote{This work has been supported by the MURST/Cofin2000 under
the project: Stellar observables of cosmological relevance.}
\begin{figure}
\plotfiddle{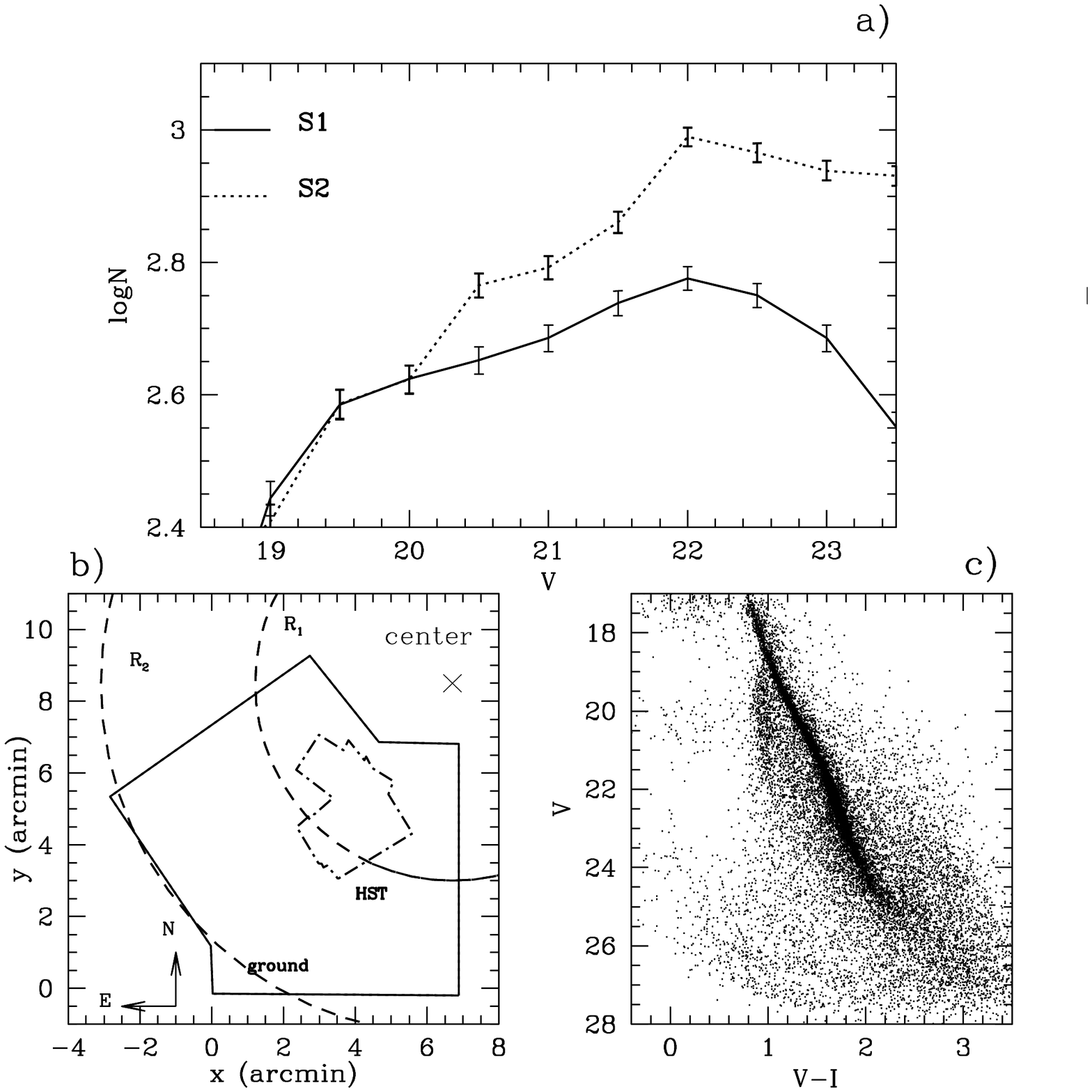}{11.6cm}{0}{60}{60}{-200/in}{-95/in}
\end{figure} 
\end{document}